\documentclass[journal]{IEEEtran}
\usepackage{cite}
\usepackage{amsmath,amssymb,amsfonts,amsthm}
\usepackage{algorithmic}
\usepackage{physics}
 \usepackage{graphicx}
\usepackage{textcomp}
\usepackage[T1]{fontenc}
\usepackage{caption}
\usepackage{subcaption}
\usepackage{epstopdf}
\usepackage{overpic}
\usepackage{array}
\usepackage[monochrome]{color}
\usepackage{url}
\usepackage{bbm}
\usepackage{enumerate}
\usepackage[shortlabels]{enumitem}

\usepackage[none]{hyphenat}
\usepackage{hyperref}
\usepackage{lipsum}

\def\BibTeX{{\rm B\kern-.05em{\sc i\kern-.025em b}\kern-.08em
    T\kern-.1667em\lower.7ex\hbox{E}\kern-.125emX}}
    
\usepackage{color}

\def\Ttran{\mbox{\tiny $\mathrm{T}$}}

\def\Real{\mathbb{R}}
\def\Complex{\mathbb{C}}

\def\diag{\mathrm{diag}}

\def\imagunit{\mathsf{j}} 

\newcommand{\vect}[1]{{\boldsymbol{#1}}}

\theoremstyle{plain}

\usepackage[margin=0.93in]{geometry}

\IEEEoverridecommandlockouts

\begin{document}

\title{\Huge{{\color{blue}On Landau's Eigenvalue Theorem for\\ Line-of-Sight MIMO Channels}}}

\author{\vspace{-0.0cm}
\IEEEauthorblockN{Andrea Pizzo \emph{Member, IEEE}, Angel Lozano, \emph{Fellow, IEEE}}
\thanks{\vspace{0.0cm}
\newline \indent 
A.~Pizzo and A.~Lozano are with Univ. Pompeu Fabra (email: \{andrea.pizzo, angel.lozano\}@upf.edu).
Work supported by the European Research Council under the H2020 Framework Programme/ERC grant agreement 694974, by the ICREA Academia program, by the European Union-NextGenerationEU, and by the Fractus-UPF Chair on Tech Transfer and 6G.
}}


\maketitle
\vspace{-0.0cm}
\begin{abstract}
An alternative derivation is provided for the degrees of freedom (DOF) formula on line-of-sight (LOS) channels via Landau's eigenvalue theorem for bandlimited signals. 
Compared to other approaches, Landau's theorem provides a general framework to compute the DOF in arbitrary environments,
this framework is herein specialized to LOS propagation. 
The development shows how the spatially bandlimited nature of the channel relates to its geometry under the paraxial approximation
that applies to most LOS settings of interest.
\end{abstract}

{\color{blue}
\begin{IEEEkeywords}\vspace{0.0cm}
Degrees of freedom, line-of-sight MIMO, paraxial approximation, Landau's eigenvalue theorem.
\end{IEEEkeywords}}

\section{Introduction}


The number of distinct waveforms able to transport information via electromagnetic waves is an inherent property of a physical channel. It is upper bounded by the number of degrees of freedom (DOF), a quantity of interest in information theory \cite{PoonDoF,Ozgur2013,Desgroseilliers2013}, optics \cite{DiFrancia,Walther,Thaning_2003,Miller,Piestun}, electromagnetism \cite{Bucci,Migliore,Janaswamy2006}, and signal processing \cite{Kennedy2007,PizzoTSP21}. 
Given the continuous nature of channels, waveforms span an
infinite-dimensional space, yet the noise allows
 for a certain error in the
 representation \cite{Migliore}. Channels are thus amenable to a discrete representation over a space of approximately DOF dimensions \cite{FranceschettiBook}.

There are various ways to compute the number of DOF in a wireless channel, say by leveraging diffraction theory \cite{DiFrancia,Walther,Thaning_2003}, by studying the eigenvalues of the Green's operator \cite{Miller,Piestun,Bucci,Desgroseilliers2013}, or by pursuing a signal-space approach \cite{PoonDoF,PizzoTSP21}. 
This paper provides an alternative derivation via Landau’s eigenvalue theorem for multidimensional bandlimited signals (or fields) \cite{FranceschettiLandau}. Analogously to time-domain waveforms of finite bandwidth, an electromagnetic channel may be regarded as spatially bandlimited due to a low-pass filtering behavior of the propagation \cite{Bucci,PoonDoF,PizzoTSP21}. In this analogy, time is replaced by space and frequency by spatial-frequency (or wavenumber) \cite{FranceschettiBook}.

{\color{blue}Originally devised for waveform channels \cite{LandauWidom}, Landau's theorem has been generalized to electromagnetic propagation \cite{FranceschettiLandau}, and lately applied to non-line-of-sight (NLOS) channels \cite{PizzoTSP21}. 
Prompted by the interest in LOS multiple-input multiple-output (MIMO) communication at high frequencies \cite{HeedongTWC}, here   
 the connection is drawn with such channels under
the paraxial approximation that holds when the propagation 
is focused about the axis connecting the two arrays \cite{GoodmanBook}.
The development builds on signal theory concepts, without relying on unconventional mathematics.
A bridge between LOS and NLOS propagation is also uncovered, with implications for MIMO communication and Nyquist reconstruction at high frequencies.}

\emph{Notation:}
$\mathcal{F}_n$ is the $n$-dimensional Fourier operator, $(\mathcal{F}_n h)(\vect{f}) = \int_{\Real^n} h(\vect{t}) e^{-\imagunit 2 \pi \vect{f}^{\Ttran} \vect{t}} \, d\vect{t} = g(\vect{f})$, whereas $\mathcal{F}_n^{-1}$ is its inverse, $(\mathcal{F}_n^{-1} g)(\vect{t}) = h(\vect{t})$,
with the shorthand notation $\mathcal{F}_1 = \mathcal{F}$ and $\mathcal{F}_1^{-1} = \mathcal{F}^{-1}$. In turn,  $(\mathbbm{1}_{R} h)(\vect{t}) = \mathbbm{1}_{R}(\vect{t}) h(\vect{t})$ with $\mathbbm{1}_{R}(\vect{t})$ the indicator function of a set $R\subset \Real^n$ while
$R_{\vect{A}}$ is the set obtained by applying any invertible linear transform $\vect{A}$ to the axes of $R$,  and $m(\mathcal{\cdot})$ is the Lebesgue measure. 
 
\begin{figure}
\centering\vspace{-0.0cm}
\includegraphics[width=.999\columnwidth]{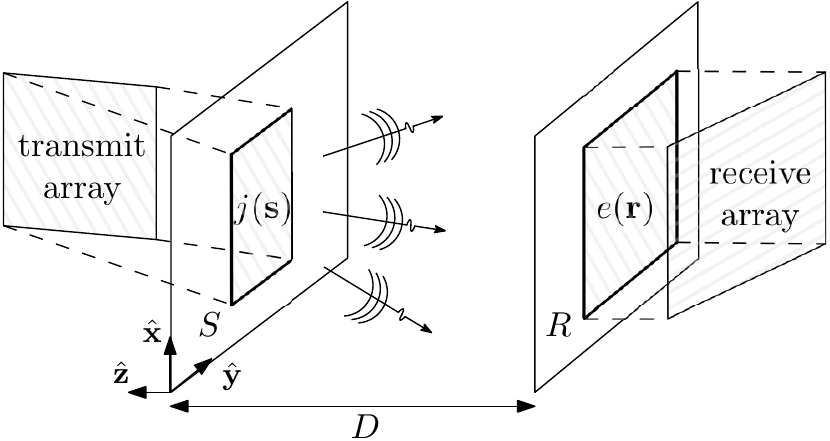} \vspace{-0.0cm}
\caption{LOS communications between continuous arrays.}\vspace{-0.0cm}
\label{fig:LOS_channel}
\end{figure}
 
%

\section{Plane-wave Representation \\ of LOS channels} \label{sec:Section_II}


Consider two $n$-dimensional continuous-space arrays ($n=1$ or $2$) communicating with scalar electromagnetic waves at wavelength $\lambda$ in a 3D free-space environment. 
We denote by $D$ the distance between the array centroids.
Capitalizing on that an arbitrary source can always be replicated by a flat source on a given plane thanks to Huygen's principle \cite{ChewBook}, we let
${S \subset \Real^n}$ and ${R \subset \Real^n}$ be the projections---respecting the respective centroids---of the source and receive arrays onto parallel planes ($n=2$) or parallel lines ($n=1$).
The normal to these planes or lines aligns with the $z$-axis, as shown in Fig.~\ref{fig:LOS_channel} for $n=2$.
The scalar electromagnetic field $e(\vect{r})$, $\vect{r} \in R$, is the image of a current density $j(\vect{s})$, $\vect{s} \in S$, through a linear channel operator $\mathcal{G}$ as
\begin{equation} \label{operator}
e(\vect{r}) = (\mathcal{G} j)(\vect{r}) =  \int_{S}  h(\vect{r},\vect{s}) j(\vect{s}) \,  d\vect{s}
\end{equation}
where $h(\vect{r},\vect{s})$ is the space-variant kernel induced by the operator, as dictated by the physical environment.
In LOS, it is found by solving \cite{ChewBook}
 \begin{equation} \label{Helmholtz}
\nabla^2  e(\vect{r}) + \left(\frac{2\pi}{\lambda}\right)^2 e(\vect{r}) = \imagunit \frac{2\pi \eta}{\lambda} j(\vect{r})
\end{equation}
with $\eta$ the impedance. 
The kernel solving \eqref{Helmholtz} is known to be given by $h(\vect{r},\vect{s}) = -\imagunit \frac{2\pi \eta}{\lambda}   \, G(\vect{r},\vect{s})$  
with 
\cite[Sec.~1.3.4]{ChewBook}  
\begin{equation} \label{Green}
G(\vect{r},\vect{s}) = \frac{e^{\imagunit 2\pi \frac{r}{\lambda}}}{4\pi r}
\end{equation}
the scalar Green's function,
where $r = \|(\vect{r}-\vect{s},-D)\|$ reveals the space-invariant nature of LOS channels \cite{PizzoIT21}.
The foregoing kernel can also be represented as the plane-wave decomposition \cite[Eq.~12]{PizzoIT21}
\begin{equation}   \label{impulse_response}
 h(\vect{r},\vect{s}) =  \frac{\eta}{2 \lambda}  \int_{\Real^n} \!\! \frac{e^{-\imagunit 2\pi \kappa_z D}}{\kappa_z} \, e^{\imagunit 2\pi \vect{k}^{\Ttran} (\vect{r} - \vect{s})}  \, d\vect{k}
\end{equation}
with
\begin{equation} \label{kappaz} 
\kappa_{z} = 
\begin{cases} \displaystyle
 \sqrt{1/\lambda^2 - \|\vect{k}\|^2}  & \quad \|\vect{k}\| \le  1/\lambda \\ \displaystyle
\imagunit  \sqrt{\|\vect{k}\|^2 - 1/\lambda^2}  & \quad  \|\vect{k}\| >  1/\lambda
\end{cases}
\end{equation}
such that $\|\vect{k}\|^2 + \kappa_{z}^2 = 1/\lambda^2$ for every wave vector\footnote{For notational convenience, the wavenumber domain in \cite[Ch.~8]{FranceschettiBook} is rescaled by $1/2\pi$, corresponding to the spatial frequency domain.} $\vect{k}$. 
Properly normalized, the real parts of $(\vect{k},\kappa_z)$ are the cosines of the angles subtended by each plane wave with the axes,
while
$ 
\lambda \|\vect{k}\|  = \sin \theta
$ 
and ${\lambda \kappa_z  = \cos \theta}$ given ${\theta \in[0,\pi/2]}$ as the plane-wave's angle with the $z$-axis.


\section{Paraxial Approximation in the Wavenumber Domain}  \label{sec:Section_III}



The \emph{paraxial approximation} applies 
 when, away from the source, propagation is focused about the axis connecting with the receiver (the $z$-axis in our case) \cite{GoodmanBook}.
It entails 
$D \gg L$ with $L$ the maximum array dimension, such that the phase and magnitude of \eqref{Green} satisfy \cite{Miller,HeedongTWC}
\begin{equation}
r \approx
\begin{cases} \label{paraxial} \displaystyle
 D + \frac{\|\vect{r} - \vect{s}\|^2}{2 D} & \text{(phase)} \\
 D & \text{(magnitude)}
\end{cases}
\end{equation}
where the phase's behavior follows from
${\sqrt{1+x} \approx 1 + \frac{x}{2}}$ for small $x = \|\vect{r} - \vect{s}\|^2/D^2$.


The paraxial approximation has its translation to the wavenumber domain. From ${|\sin \theta | \ll 1}$, it follows that ${\|\vect{k}\| \ll 1/\lambda}$; then, \eqref{kappaz} satisfies \cite{GoodmanBook}
 \begin{equation} \label{paraxial_wavenumber} 
 \kappa_{z} \approx
\begin{cases}  \displaystyle
  {1}/{\lambda} - \frac{\lambda\|\vect{k}\|^2}{2} & \text{(phase)} \\ \displaystyle
 {1}/{\lambda} & \text{(magnitude)}.
\end{cases}
\end{equation}

It is shown in Appendix~\ref{app:paraxial_integral} that, under the paraxial approximation, (\ref{operator}) reduces to
\begin{equation} \label{input_output_Fresnel}
e(\vect{r}) = (\widehat{\mathcal{G}} j)(\vect{r}) = \int_{S} e^{\imagunit 2\pi \vect{s}^{\Ttran} \! \frac{\vect{r}}{\lambda D}} j(\vect{s}) \,  d\vect{s}
\end{equation}
for $\vect{r}\in R$.
The uniform scaling transform $\vect{A} = \frac{1}{\lambda D} \vect{I}_n$ of the receiver's axes would yield, equivalently,
\begin{equation} \label{input_output_Fresnel_Fourier}
e(\vect{r})  = (\mathcal{H} j)(\vect{r}) =    \int_{S}  j(\vect{s}) \, e^{\imagunit \vect{s}^{\Ttran} \vect{r}}  \,  d\vect{s}
\end{equation}
for $\vect{r}\in R_{\vect{A}}$.
Hence, 
paraxial LOS channels amount to an $n$-dimensional inverse Fourier transform of the source density 
returning the received field \cite{GoodmanBook}. 
The limitation of the source support that a transmit array imposes
corresponds to a low-pass filtering operation, revealing the \emph{spatially bandlimited} nature of electromagnetic fields \cite{Bucci,PoonDoF,PizzoTSP21}.
With respect to the classical definition of a bandlimited signal in the frequency domain, here, the notion applies in the wavenumber domain \cite[Ch.~8]{FranceschettiBook}.


\section{Kolmogorov Space Dimensionality}  \label{sec:Section_IV}


{\color{blue}Due to conservation of energy, $e(\vect{r})$ belongs to the Hilbert space $V$ of square-integrable functions. 
This space is equipped with the norm $\|e\| =  ( \int_{\Real^n} |e(\vect{r})|^2 \, d\vect{r} )^{1/2}$ with $e(\vect{r})$ characterized by an infinite number of basis functions.
The DOF provide a measure of the {effective dimensionality} of $V$, i.e., the minimum number $N$ of basis functions needed to represent every element of $V$ up to some accuracy.}
The degree of approximation of $V$ by an $N$-dimensional subspace $V_N$ is measured by the Kolmogorov $N$-width \cite[Ch.~3.2]{FranceschettiBook}
\begin{equation} \label{n_width}
d_N(V) = \inf_{{\rm dim}(V_N)=N}  D_{V_N}(V)
\end{equation}
with $D_{V_N}(V) =  \sup_{e\in V}  \inf_{e_N\in V_N} \|e - e_N\|$
 the deviation between $V$ and $V_N$ according to a min-max criterion.
The $N$-width in \eqref{n_width} is the smallest such deviation over all subspaces of dimension $N$. 
The DOF in $V$ at any level of accuracy ${0 < \sigma < 1}$ is then
 \begin{equation} \label{dof}
{\sf DOF}_\sigma = \min\{N : d_N(V) \le \sigma\}
\end{equation}
whose existence is ensured by the spectral theorem for self-adjoint operators.
Precisely, for any
Hilbert-Schmidt operator $\mathcal{H}$, we have that  \cite[Eq. 3.56]{FranceschettiBook}
\begin{equation} \label{dof_lambda}
{\sf DOF}_\sigma = \min\{N : \lambda_N \le \sigma\}
\end{equation}
where $\lambda_N$ is the $N$th smallest eigenvalue of 
$\mathcal{H} \mathcal{H}^*$ (composition of $\mathcal{H}$ with its adjoint $\mathcal{H}^*$).
The discrete counterpart is the spectral theorem for Hermitian matrices, with $\lambda_N$ the $N$th smallest eigenvalue of $\vect{H} \vect{H}^*$.

\section{DOF} \label{sec:Section_V}
\subsection{Bandlimited Waveforms}

As the observation interval $T$ increases, a waveform concentrates within a bandwidth $B$ (in Hz), with the maximum simultaneous concentration in time and frequency dictated by the uncertainty principle \cite[Ch.~2]{FranceschettiBook}.
This behavior is specified by \cite{LandauWidom}
\begin{equation} \label{eigenvalue_operator}
\mathcal{T}_T \mathcal{B}_B \mathcal{T}_T  \phi_i(t) =  \lambda_i \phi_i(t)
\end{equation}
where 
$\mathcal{T}_T = \mathbbm{1}_{|t| \le T/2}$ and $\mathcal{B}_B = \mathcal{F}^{-1} \mathbbm{1}_{|f| \le B} \mathcal{F}$ correspond to time-limiting to $T/2$ and frequency-limiting to $B$ \cite[Ch.~3.4.1]{FranceschettiBook}. 
Rewriting \eqref{eigenvalue_operator} as $\mathcal{H} \mathcal{H}^*$ with 
\begin{equation} \label{H}
\mathcal{H} = \mathcal{T}_T  \mathcal{F}^{-1} \mathbbm{1}_{|f| \le B}   \qquad
\mathcal{H}^* = \mathbbm{1}_{|f| \le B} \mathcal{F} \mathcal{T}_T,
\end{equation}
the spectral theorem yields an eigensolution. 
Specifically, for any $\sigma$, \eqref{dof_lambda} is obtained by spectral concentration after 
letting $T$ grow while keeping $B$ fixed, 
giving \cite[Eq.~2]{LandauWidom} 
\begin{equation}  \label{dof_time}
 {\sf DOF}_\sigma  \!=\!
{\sf DOF} +  \frac{1}{\pi^2} \log \! \left(\frac{1-\sigma}{\sigma} \right) \log T + o(\log T )
\end{equation}
where 
\begin{equation} \label{N0_time}
{\sf DOF}  = 2BT.
\end{equation}
By symmetry, \eqref{dof_time} can also be obtained from an operator $\mathcal{B}_B \mathcal{T}_T \mathcal{B}_B$, scaling the frequency axis by $B$ and letting $B$ grow while keeping $T$ fixed.

\subsection{Spatially Bandlimited Fields}

Generalization 
to multidimensional signals (or fields) is achieved by replacing time with space, and frequency with wavenumber. The concentration of a spatially bandlimited field of wavenumber support $\vect{k} \in K \subset \Real^n$ observed on a region $R_{\vect{A}} \subset \Real^n$ is ruled by \cite[Eq.~10]{FranceschettiLandau} \cite[Ch.~3.5]{FranceschettiBook}
\begin{equation} \label{eigenvalue_operator_2D}
\mathcal{T}_{R_{\vect{A}}} \mathcal{B}_{K} \mathcal{T}_{R_{\vect{A}}}  \phi_i(\vect{r}) =  \lambda_i \phi_i(\vect{r})
\end{equation}
where 
$\mathcal{T}_{R_{\vect{A}}} = \mathbbm{1}_{R_{\vect{A}}}$ and $\mathcal{B}_{K} = \mathcal{F}_n^{-1} \mathbbm{1}_{K} \mathcal{F}_n$ correspond to space-limiting to $R_{\vect{A}}$ and wavenumber-limiting to $K$. 
Rewriting \eqref{eigenvalue_operator_2D} as $\mathcal{H} \mathcal{H}^*$ with
\begin{equation} \label{H}
\mathcal{H} = \mathcal{T}_{R_{\vect{A}}} \mathcal{F}_n^{-1} \mathbbm{1}_{K}  \qquad
\mathcal{H}^* = \mathbbm{1}_{K} \mathcal{F}_n \mathcal{T}_{R_{\vect{A}}},
\end{equation}
an eigensolution of \eqref{eigenvalue_operator_2D} is obtained as \cite{FranceschettiLandau}, 
\begin{align} \label{dof_time_2D} 
{\sf DOF}_\sigma  & = 
{\sf DOF} +  \frac{1}{\pi^2} \log \! \left(\frac{1-\sigma}{\sigma} \right) \log \det(\vect{A}) \\ & \quad 
+ o(\log \det(\vect{A})) \nonumber
\end{align}
where
\begin{equation}  \label{N0_space}
{\sf DOF}  =  m(K) m(R_{\vect{A}}).
\end{equation}
Spectral concentration arises as $R_{\vect{A}}$ varies over $\vect{A} R$ with fixed $R$ and growing $\det(\vect{A})$, while $K$ is fixed \cite[Ch.~3.5.4]{FranceschettiBook}.
By symmetry, \eqref{dof_time_2D} is also obtainable from an operator $\mathcal{B}_{K} \mathcal{T}_{R_{\vect{A}}} \mathcal{B}_{K}$ after scaling the wavenumber domain by a growing $\det(\vect{A})$ while $R$ and $K$ are fixed.

From (\ref{N0_space}), we can recover \eqref{N0_time} by setting $R = \{|t|\le 1/2\}$ and $K = \{|f |\le B\}$ while turning $\vect{A}$ into $T$.

\subsection{Paraxial LOS Channels}  \label{sec:paraxial}

Owing to the Fourier relationship between source current and receive field in (\ref{input_output_Fresnel_Fourier}), restricting the source to $S$
is tantamount to limiting the wavenumber to
 $K = S$ at the receiver. In turn, the receiver region is $R_{\vect{A}}$ 
with $\vect{A} = \frac{1}{\lambda D} \vect{I}_n$. 
The DOF are then an instance of \eqref{N0_space}, precisely
\begin{equation} \label{DOF} 
{\sf DOF} = m(S) m(R_{\vect{A}}) = \frac{ m(S) m(R) }{(\lambda D)^n} 
\end{equation}
given ${m(R_{\vect{A}}) = \det(\vect{A}) m(R)}$.
{\color{blue}Spectral concentration is achieved with ${R}$ and ${S}$ fixed while $\lambda D$ shrinks, whereby the receive array
becomes magnified from the vantage of the source and
 the spatial resolution sharpens.} 
Although such concentration could potentially be squeezed by the paraxial approximation, which requires $D$ to be large, it is shown in Sec.~\ref{sec:asymptotic}
that such squeeze is rather inconsequential. 


Symmetry may be leveraged to alternatively obtain \eqref{DOF} via 
$\mathcal{B}_K \mathcal{T}_{R_{\vect{A}}} \mathcal{B}_K$.
This amounts to swapping source and receiver, with the same result due to reciprocity \cite{PizzoIT21}.

\begin{figure}
\centering\vspace{-0.0cm}
\includegraphics[width=0.999\columnwidth]{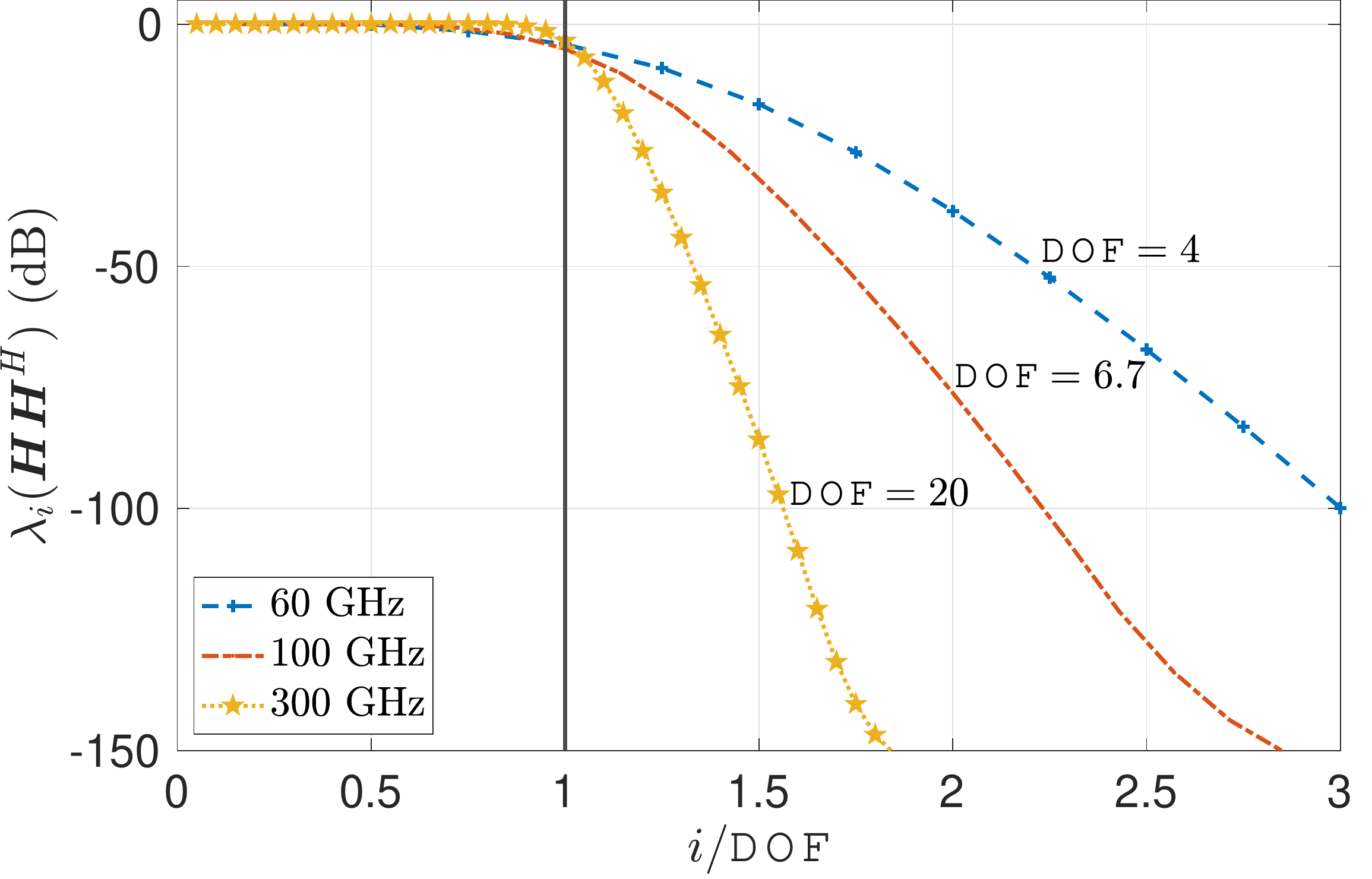} \vspace{-0.4cm}
\caption{Normalized sorted eigenvalues of $\vect{H} \vect{H}^* \in \Complex^{N \times N}$ at various frequencies for ULAs with $L = 0.2$~m and $D = 10 L$. Smooth curves are obtained by having $N \gg {\sf DOF}$.} 
\vspace{-0.1cm}
\label{fig:dof_ula}
\end{figure}

{\color{blue}Let $\vect{H} \in \Complex^{N \times N}$ be the MIMO channel matrix obtained by sampling two 1D continuous arrays of size $L = 0.2$~m. 
The eigenvalues of $\vect{H} \vect{H}^{*}$ are plotted in Fig.~\ref{fig:dof_ula}, sorted and normalized by ${\sf DOF} = {L}^2/(\lambda {D})$. The frequency is $\{60,100,300\}$~GHz and ${D} = 10 {L}$, complying with the paraxial approximation. Expectedly, the eigenvalues polarize into two levels as the carrier frequency increases due to spectral concentration. }

\section{LOS and NLOS propagation}

\subsection{DOF}

NLOS channels are specified by $h(\vect{r},\vect{s})$, space-variant because of multipath propagation introducing a separate dependence on the source and receiver locations \cite{PizzoIT21}. 
For 1D arrays of dimensions $L_{\rm s}$ and $L_{\rm r}$, under isotropic scattering \cite[Eq.~27]{PoonDoF}
\begin{equation} \label{DOF_1D}
{\sf DOF} =  {\min \! \left(\frac{L_{\rm s}}{\lambda/2}, \frac{L_{\rm r}}{\lambda/2}\right)} .
\end{equation}
In complete generality, with $n$-dimensional arrays and wavenumber supports ${K_{\rm s} \subset \Real^n}$ and ${K_{\rm r} \subset \Real^n}$,
\begin{equation} \label{DOF_NLOS}
{\sf DOF}  = \min\!\Big( m(K_{\rm r}) m(C_{\vect{A}_{\rm r}}),  \, m(K_{\rm s}) m(C_{\vect{A}_{\rm s}})\Big)
\end{equation}
where the source and receiver have been expressed as linear transformations of a unit-measure set $C \subset \Real^n$ with transform matrices $\vect{A}_{\rm s}$ and $\vect{A}_{\rm r}$, i.e., $S = C_{\vect{A}_{\rm s}}$ and $R = C_{\vect{A}_{\rm r}}$.
Spectral concentration occurs when $\min( \det (\vect{A}_{\rm s}),\det (\vect{A}_{\rm r}))$ grows while $K_{\rm s}$ and $K_{\rm r}$ are fixed.

To see how \eqref{DOF_NLOS} specializes to \eqref{DOF_1D}, it suffices to let $C = \{|x|\le 1/2\}$ and $K_{\rm s} = K_{\rm r} = \{|\kappa_x|\le 1/\lambda\}$ while $\vect{A}_{\rm s}$ and $\vect{A}_{\rm r}$ become the scalars $L_{\rm s}$ and $L_{\rm r}$.

{\color{blue}From  \eqref{DOF_NLOS}, we can also geometrically recover the result for paraxial LOS channels in \eqref{DOF}. 
The solid angle subtended by the source at the receiver is $m(S)/D^n$. From the paraxial approximation, $\sin(\theta) \approx \theta$
while 
$\lambda \|\vect{k}\| = \sin \theta$, 
hence
${m(K_{\rm r}) = m(S)/(\lambda D)^n}$. Due to reciprocity, ${m(K_{\rm s}) = m(R)/(\lambda D)^n}$ from the vantage of the source. Plugging these results into \eqref{DOF_NLOS} yields \eqref{DOF}.

There are two terms in \eqref{DOF_NLOS}, modeling the separate scattering at both ends of the link and the ensuing space variance of NLOS channels \cite{PizzoIT21}. In contrast, there is only a term in \eqref{DOF}, as LOS propagation puts source and receiver in one-to-one correspondence, leading to space invariance.

Also noteworthy is that,
while in LOS channels $S$ and $R$ are the projections of the source and receive arrays, in NLOS channels, these are the actual array apertures as the array orientations are embedded into the angular selectivity of the local scattering.}

\subsection{Asymptotic Regimes} \label{sec:asymptotic}

Another difference between \eqref{DOF} and \eqref{DOF_NLOS} is in their regimes of relevance, where eigenvalues polarize into two levels (see Fig.~\ref{fig:dof_ula}) and asymptotic results can be leveraged \cite{HeedongTWC}.
To bring out the key concept, consider rectangular arrays of dimensions $L_{{\rm s},i}$ and $L_{{\rm r},i}$, $i=1, \ldots, n$, which arise from the transformation of a unitary square 
$C$ by
\begin{align}
\vect{A}_{\rm s}  & = \diag(\{L_{{\rm s},i}\}_{i=1}^n) \\
\vect{A}_{\rm r} &  = \diag(\{L_{{\rm r},i}\}_{i=1}^n).
\end{align}
The spectra of NLOS channels concentrate for 
\begin{equation}
 \min\! \left(\prod_{i=1}^n \frac{L_{{\rm s},i}}{\lambda},\prod_{i=1}^n \frac{L_{{\rm r},i}}{\lambda}\right) \gg 1 ,
\end{equation}
implying electrically large arrays. 
Alternatively, 
paraxial LOS channels require
\begin{equation} \label{asymp}
 \sqrt[n]{\prod_{i=1}^n \frac{L_{{\rm s},i}}{\lambda} \prod_{i=1}^n \frac{L_{{\rm r},i}}{\lambda}} \gg \frac{D}{\lambda} \gg \max_{i=1,n}\!\left(\frac{L_{{\rm s},i}}{\lambda},\frac{L_{{\rm r},i}}{\lambda}\right)
\end{equation}
where the first inequality ensures spectral concentration in \eqref{DOF} and the second one embodies the paraxial approximation.
(As a by-product of the first inequality, $D$ is also incompatible with planar wavefronts.)

Welcomely, (\ref{asymp}) delimits a broad range of validity for the developed theory.
With squared arrays of size $L$, setting $D= 10 L$ as a reasonable concretization of the second inequality,
the first one yields $D/\lambda \gg 100$; at $100$~GHz, this amounts to $D \gg 0.3$~m. Rescaling one axis by $\beta \ge 1$ and the other one by $1/\beta$, to keep the array apertures fixed while altering their aspect ratio, setting $D= 10 \beta L$ yields $D/\lambda \gg 100 \beta^2$; at $100$~GHz with $\beta=4$, this gives $D \gg 4.8$~m.



\subsection{Nyquist Sampling}

The DOF per spatial unit correspond to the sampling density $\mu$ (in samples/m$^n$) needed for reconstruction \cite{PizzoTSP21}, extending the classical notion of Nyquist rate (samples/s) to $n$-dimensional fields. 

Recalling \eqref{DOF_NLOS} for NLOS channels, at the receiver
\begin{equation} \label{Nyq_NLOS}
\mu_{\rm r} = \frac{{\sf DOF}}{m(R)} = m(K_{\rm r}).
\end{equation}
This is highest under isotropic scattering, when $K_{\rm r}$ is an $n$-dimensional disk of unit radius \cite{PizzoIT21,PizzoTSP21}
leading to $\lambda/2$-sampling and to hexagonal sampling with density $\pi/\lambda^2$, respectively when $n=1$ and $n=2$ \cite{PizzoTSP21}.
Scattering selectivity shrinks $m(K_{\rm r})$, rendering the sampling sparser.

In turn, recalling  \eqref{DOF} for LOS channels, at the receiver
\begin{equation} \label{Nyq_LOS}
\mu_{\rm r} = \frac{{\sf DOF}}{m(R)} = \frac{m(S)}{(\lambda D)^n}
\end{equation}
which 
depends on sheer geometry (source dimension, wavelength, and range),
rather than on the scattering selectivity.
An inspection of \eqref{Nyq_LOS} also reveals that LOS channels 
can be reconstructed more efficiently due to a lower DOF density.
For instance, $\lambda/2$-sampling with ${n=1}$ implies ${D = L_{\rm s}/2}$, which is unfeasible under the paraxial approximation.

\subsection{Rayleigh Spacing for LOS Channels}

Consider 1D arrays and let $N_{\rm s}$ and $N_{\rm r}$ be the transmit and receive antenna numbers, with uniform spacings $\delta_{\rm s} =1/\mu_{\rm s}$ and $\delta_{\rm r} = 1/\mu_{\rm r}$ (in {m/sample}); these are reciprocals of the Nyquist densities. For LOS channels, from \eqref{Nyq_LOS} and
$L_{\rm s} = (N_{\rm s}-1) \delta_{\rm s}$, we obtain $\delta_{\rm s} \delta_{\rm r} = \frac{\lambda D}{N_{\rm s}-1}$.
From reciprocity, we further infer $\delta_{\rm s} \delta_{\rm r} = \frac{\lambda D}{N_{\rm r}-1}$.
To prevent aliasing in the wavenumber domain, the antenna spacings must yield the largest spectrum separation, namely
\begin{equation} \label{Ray_LOS_tot}
 \delta_{\rm s} \delta_{\rm r} = \frac{\lambda D}{N_{\sf max}-1}
\end{equation}
with $N_{\sf max} = \max(N_{\rm r},N_{\rm s})$. 

{\color{blue}The so-called Rayleigh spacings $d_{\rm s}$ and $d_{\rm r}$, which enable full DOF exploitation and are therefore optimum at a high signal-to-noise ratio,
 satisfy \cite[Eq.~10]{HeedongTWC}
\begin{equation} \label{Rayleigh}
d_{\rm s} d_{\rm r} = \frac{\lambda D}{N_{\sf max}},
\end{equation}
which coincides with \eqref{Ray_LOS_tot} when $N_{\sf max} \gg 1$, i.e., when Nyquist sampling attains perfect reconstruction \cite{PizzoTSP21}.}

\section{Conclusion} \label{sec:Section_VI}

{\color{blue}The paraxial approximation endows LOS channels with a bandlimited nature in the wavenumber domain, a nature from which
the DOF formula can be obtained via Landau's eigenvalue theorem  \cite{LandauWidom,FranceschettiLandau}.
As in NLOS channels \cite{PoonDoF,Ozgur2013}, the ensuing DOF are determined by the size and geometry of the arrays and the angular selectivity of the environment. LOS channels are inherently geometrical \cite{HeedongTWC}, with 
the angular selectivity dictated by the solid angle subtended by the source at the receiver. 
Three physical effects play a role:
\emph{zooming}, inversely proportional to the wavelength, \emph{skewing}, function of the relative array orientations,  
and \emph{magnification} as the communication range shrinks.}
 


\appendices

\section{} \label{app:paraxial_integral}

Plugging (\ref{paraxial_wavenumber}) into the plane-wave representation in \eqref{impulse_response},
\begin{equation}   \label{impulse_response_paraxial}
\widehat h(\vect{r},\vect{s}) =  \frac{\eta}{2 \lambda} e^{-\imagunit 2\pi \frac{D}{\lambda}} \int_{\Real^n} \!\! e^{\imagunit \pi  \lambda D \|\vect{k}\|^2}  e^{\imagunit 2\pi \vect{k}^{\Ttran} (\vect{r}-\vect{s})} \, d\vect{k} ,
\end{equation}
from which, removing unwanted constants, 
\begin{equation}   \label{impulse_response_paraxial_2}
 \widehat h(\vect{x}) =  \int_{\Real^n} e^{\imagunit \pi \lambda D \|\vect{k}\|^2}  e^{\imagunit  2\pi \vect{k}^{\Ttran} \vect{x}} \, d\vect{k}
\end{equation}
where 
$\vect{x} = \vect{r}-\vect{s}$. 
Eq. \eqref{impulse_response_paraxial_2} can be computed independently along each axis, e.g., along the $x$-axis,
\begin{align}    \label{separable}
\widehat h(x) & =  \int_{-\infty}^{\infty} e^{\imagunit \pi \lambda D \kappa_x^2} \, e^{\imagunit 2\pi \kappa_x x}  \, d\kappa_x.
\end{align}
We recall that \cite[Eq.~7.4.6]{AbramowitzStegun},
\begin{equation} \label{Fourier_result_inverse}
\int_{-\infty}^\infty e^{- \frac{\pi^2}{a} f^2} e^{\imagunit 2\pi f t} \, df =  \sqrt{\frac{a}{\pi}} \, e^{-a t^2}
\end{equation}
for any $a \in \Complex$ with $\Re(a)>0$.
Contrasting \eqref{Fourier_result_inverse} with \eqref{separable}, we set ${a= \frac{\imagunit \pi}{\lambda D}}$
to obtain $\widehat h(x) =   e^{-\frac{\imagunit \pi}{\lambda D} x^2}$
where all known constants have been omitted. The condition $\Re(a)>0$ maps to a lossy medium.
Reintroducing vector notation to account for $n$-dimensional arrays and omitting all known constants,
\begin{equation} \label{Fourier_final}
\widehat h(\vect{r},\vect{s}) =  e^{-\frac{\imagunit \pi}{\lambda D} \|\vect{r}-\vect{s}\|^2}.
\end{equation}
Expanding and rearranging the quadratic terms in \eqref{Fourier_final}, we obtain \cite{HeedongTWC}
\begin{equation} \label{Fourier_final_expanded}
\widehat h(\vect{r},\vect{s}) =  \phi(\vect{r}) \, e^{\imagunit \frac{2\pi}{\lambda D} \vect{s}^{\Ttran} \vect{r}} \, \phi(\vect{s})
\end{equation}
where ${\phi(\vect{x}) = \exp(-\imagunit \frac{\pi}{\lambda D} \|\vect{x}\|^2)}$.
The channel kernel entails two separable quadratic phase shifts and a cross phase shift that depends on the relative source and receive locations.
With the geometry known at each end of the link, $\phi(\vect{s})$ and $\phi(\vect{r})$ can be compensated for, 
 yielding \eqref{input_output_Fresnel}.

\section*{Acknowledgment}

Comments and suggestions by Heedong Do, from POSTECH, are gratefully acknowledged.

\bibliographystyle{IEEEbib}
\bibliography{refs}

\end{document}